\documentstyle[epsfig,12pt,amssym,here]{article}
\def\smd2{a diagonally off-set two layer SMD }
\def\smd{SMD-dos}
\def\g{\gamma}
\def\pizero{\pi^{0}}
\def\rejpi{$\pi^0$ rejection efficiency }
\def\rejpis{$\pi^0$ rejection efficiencies }
\begin{document}

\begin{titlepage}
\bigskip
%\bigskip
\begin{tabbing}
\` {\large TAUP 2369-96} \\
\` September 25, 1996 \\
%\` \today\\
\end{tabbing}
\bigskip
\begin{center}{\Large\bf  Separation of
\mbox{$\gamma$/$\pi^{0}$}Showers at High Energies}
\end{center}
\begin{center}{\Large\bf
J. Grunhaus and S. Kananov\\
}\end{center}\bigskip
\begin{center}{\large School of Physics and Astronomy,
Raymond and Beverly Sackler Faculty of Exact Sciences,
Tel Aviv University, \mbox{Tel Aviv 69978,} ISRAEL
}\end{center}\bigskip
\bigskip\bigskip
\begin{center}{\it Submitted to Nuclear Instruments and Methods.\\
}\end{center}\bigskip
\begin{center}{\Large\bf Abstract}\end{center}
\normalsize
We have designed  and carried out simulation studies of a two layer
Shower Maximum Detector diagonally off-set (SMD-dos) optimized
for the separation of $\pi^0$ showers from $\gamma$  showers in
the 30 to 150 GeV energy range.
For 90$\%$ $\g$ acceptance the SMD-dos yields
\rejpis  of  92$\pm$4 $\%$,
87$\pm$4 $\%$ and 32$\pm$2 $\%$, respectively, for
30, 50 and 150 GeV incident energies. We find that the SMD-dos is
superior to  a
conventional geometry single-layer or mutiple-layer shower maximum
detector (SMD), of equal granularity, by an average factor
of $\sim$ 1.5 over the 50 to 150 GeV energy range.
We also find, the unexpected result, that the  SMD-dos gives better
$\pi^0$ rejection, for the same number of channels, than a SMD.
At hadron - hadron colliders the signature of choice for the
detection of the Higgs particle, in the mass range of 120 to 160 GeV,
is via the decay \mbox{$H \to \gamma \gamma$}. The addition of a
SMD-dos to the planned detectors at the LHC would significantly
reduce the background to the $\gamma$ signal coming from prolific
$\pi^0$ production.
\end{titlepage}
 
\section{Introduction}
 
The challenging task of separating $\pi^0$ initiated showers
from $\gamma$ initiated  showers, at high incident energies, has over
the past few years attracted a lot of interest and motivated extensive
R and D work \cite{seez,shev,rocky,telav,kek,spageti,bonam}.
The
scheduled building of the LHC machine has added new interest to the
\mbox{$\gamma$/$\pi^{0}$} separation problem.
The main motivation for
building  the LHC machine is the quest for the Higgs particle.
The Standard Model has weathered extremely well strenuous experimental
scrutiny during the past decades; however, it still lacks experimental
support for the predicted mass generating Higgs mechanism. The only
viable explanation of the masses of the elementary particles in the
framework of the Standard Model is the Higgs mechanism.
The two main collaborations approved for operation at  the LHC, ATLAS
\cite{ATLAS} and CMS\cite{CMS} state in their Technical Reports
the paramount importance of the searches for the Higgs particle at
the LHC machine.
The upgrading of LEP to LEP2 will allow the search for
the Higgs particle up to a mass of about 100 GeV.
Searches for the Higgs particle at higher masses will have to wait
for the operation of the LHC machine.

We have designed  and carried out simulation studies of a two layer
Shower Maximum Detector diagonally off-set (SMD-dos) optimized
for the separation of $\pi^0$ showers from $\gamma$  showers in
the 30 to 150 GeV energy range.
 A comparison of the SMD-dos with a
conventional geometry single-layer and mutiple-layer shower maximum
detector (SMD),
having the same granularity, shows that the SMD-dos yields
higher \rejpi in the energy range studied.
 
We have also carried out a comparison of the SMD-dos versus a single
layer SMD with  both detectors having the same number of
channels. We find the unexpected result that the SMD-dos yields better
 \rejpi than the SMD.
 
The electromagnetic showers were generated and studied using the
Monte Carlo program GEANT 3.15 implementing
 a 100 KeV cut for $\gamma$'s and electrons.
The shower maximum detector, either a  SMD-dos or a SMD, was embedded
inside a  E.M. calorimeter, consisting of a mixture of Pb and
scintillator (CH) having a radiation length of 0.8 cm.
The performance of the SMD-dos was optimized by carrying
 out extensive
Monte Carlo studies of the \mbox{$\gamma$/$\pi^{0}$}
separation efficiency
as a function of the following design parameters:
number of scintillator layers, thickness of scintillator layer,
granularity (cell size) and position of the SMD-dos or SMD inside
the E.M. calorimeter. Typically, samples of 500 showers were generated
for each configuration of design parameters and incident energy.
The distance of the E.M. calorimeter from the interaction region was
kept fixed at 150 cm, the design distance of the CMS detector approved
to operate at the LHC machine, and the showers were generated
with the shower initiating particle, a $\gamma$ or $\pi^{0}$,
normally incident on the E.M. calorimeter.
We have
 studied the stand-alone capabilities of the SMD-dos; however,
since the algorithms developed assume that the energy of the
shower, in question, is known, the SMD-dos must be used
with an E.M. calorimeter to determine the total energy of the shower.

At hadron - hadron colliders the signature of choice for the
detection of the Higgs particle, in the mass range of 120 to 160 GeV,
is via the decay $H \to \gamma \gamma$ ~\cite{higs}. These searches
will encounter high $\gamma$ backgrounds
coming from prolific $\pi^0$ production which decay into two $\gamma$'s.
The attainment of good  \mbox{$\gamma$/$\pi^{0}$}  separation is
of crucial importance to the success of the proposed searches.
The incorporation of a SMD-dos into the planned detectors at the LHC
would appreciably reduce the $\gamma$ background.
 
\section{The Geometry of the Diagonally Off-set
Shower Maximum Detector}
 
\begin{figure}[H]
\centering
\epsfig{file=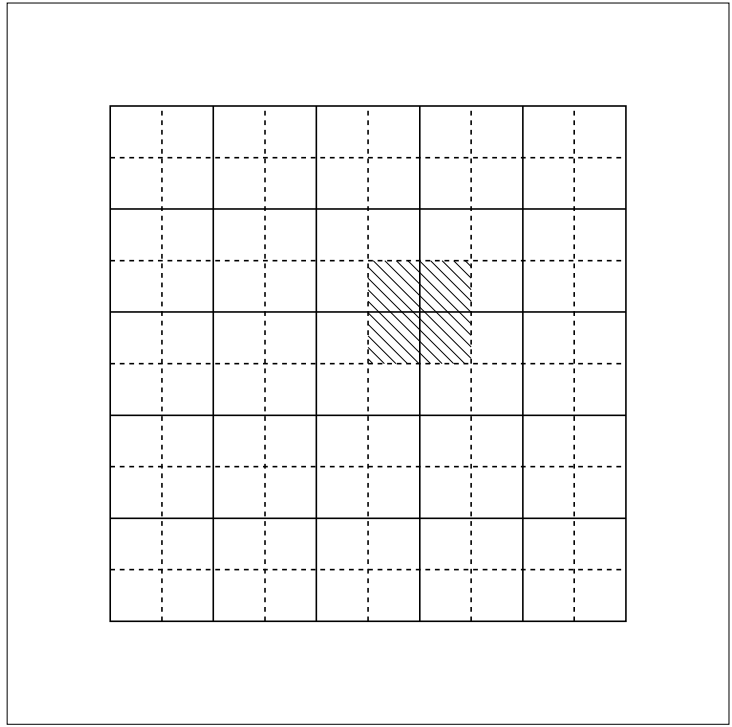,height=12.cm}
\caption{ Layout of the two layers of the SMD-dos. The cells of
the front layer are shown in solid lines and the cells of the back
layer are shown in dashed lines. One back layer cell is highlighted
to show the four quarter cells in the front layer which
overlap it.}
\label{grid1}
\end{figure}

The geometry of the \smd,
  shown in Figure~\ref{grid1},  consists of 2
identical scintillator layers, placed one behind the other.  Each layer
is subdivided into square cells which are individually read out and the
 energy deposited in each cell is recorded.
The 2 layers are diagonally
off-set with respect to each other by half a cell, so that
the intersection  of any four adjacent cells in one layer
corresponds to the center of a cell in the other layer.
 Studies of the
\mbox{$\gamma$/$\pi^{0}$} separation efficiency
 of granular
detectors show that the separation efficiency decreases as
a function of the distance of the shower center from the nearest
intersection.
Hence increasing the number of intersections per unit area enhances
the separation efficiency.
The area of each
cell, in the front layer, overlaps symmetrically one quarter of the
area of each of four adjacent cells in the back layer. And likewise
the area of each cell, in the back layer, overlaps symmetrically
one quarter of the area of each of four adjacent cells in the front
layer. Therefore, the energy collected by a particular cell is
subdivided unto 4 cells giving a substantial increase in granularity.
In the singular case that all
  the energy deposited by a
shower is confined, in a particular layer, to one cell it is
impossible with a conventional one-layer or multiple-layer SMD
 to determine whether the shower was initiated by a
 $\gamma$ or $\pi^0$.
However, the unique geometry of the SMD-dos, which subdivides
 the energy collected by a particular cell, in one layer, unto four
cells in the other layer, does furnish some degree of
\mbox{$\gamma$/$\pi^{0}$} separation efficiency even in this
singular case.

\section{Definition of Variables Used in the Algorithms}
 
The energy and position
information collected, per event, by the scintillator
cells  is used to construct 10 variables.
\begin{figure}[H]
\centering
\epsfig{file=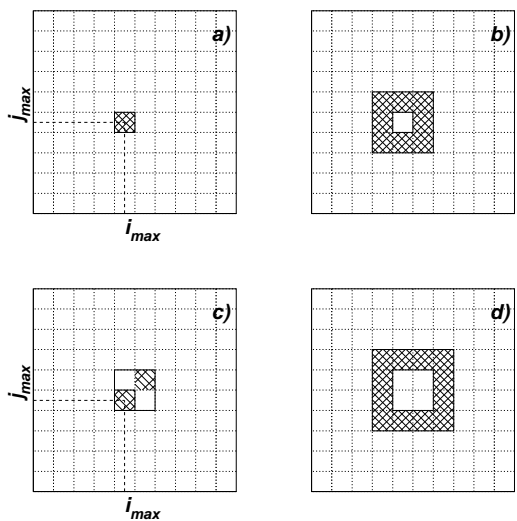,height=8.cm}
\caption{Display of cell configurations used in the determination of
 $E_{max}$, $E_{4}$, $E_{8}$ and $E_{12}$.}
\label{grid2}
\end{figure}
These  variables  embody
the experimental information which goes into the
\mbox{$\gamma$/$\pi^{0}$} separation algorithms. The information
collected in the first layer is used to construct 5 variables and
likewise the information from the second layer is used to construct
the other 5 variables. The detailed description and construction of these
10 variables is detailed in this section.
In Figure~\ref{grid2} are displayed the cell configurations used for
the determination of $E_{max}$, $E_{4}$, $E_{8}$ and $E_{12}$.
 
\begin{itemize}
\item   $<r>$, the average energy weighted radius of the shower.
The determination of $<r>$ from the data collected by the cells in
the layer is discussed in the following paragraph.
\item  $E_{max}$, the
maximum energy deposited in a single cell, in the layer.
 See Fig. 2a.
\item   $E_{8}$, the sum of energies deposited
in the eight cells which surround the cell with the maximum
energy.  $E_{8}$ is normalized to $E_{total}$.
 See Fig. 2b.
\item   $E_{4}$, the sum of energies deposited in a 4 cell square
which includes the cells with the maximum and next to maximum energy.
 $E_{4}$ is normalized to  $E_{total}$. See Fig. 2c.
\item   $E_{12}$, the sum of energies deposited in the 12 cells which
surround the  4 cell square described in the preceding item.
$E_{12}$ is normalized to  $E_{total}$.  See Fig. 2d.
\end{itemize}

The energy information gathered by each  cell is read out separately and
is labelled,  $E_{ij}$, where the x and y coordinates of the
particular cell, in the two dimensional layer, are denoted,
respectively, by the  $i$ and $j$ subscripts. The  average energy
weighted radius, $<r>$, is calculated for each layer separately.
 First the coordinates of the energy weighted center of the shower,
 $ x_{0}$ and $y_{0}$, are calculated as follows:
\mbox{ $x_{0} = \sum_{i,j} E_{ij} x_i/\sum_{i,j} E_{ij}$} and
\mbox{ $y_{0} = \sum_{i,j} E_{ij} y_i/\sum_{i,j} E_{ij}$},
where the $i$ and $j$
summations are carried out over all the cells that have
collected energy and for \(i < j\), to avoid double counting.
 Using
the  $ x_{0}$ and $y_{0}$ values,
we determine the distances
$r_{ij}$ between the center of the ${ij}$ cell and the shower
energy weighted center,
\mbox{ $r_{ij} =\sqrt{(x_{0}-x_i)^2 + (y_{0}-y_j)^2}$}.
The average energy weighted radius of the shower,
$<r>$, is determined as follows:
 
$$<r> =\frac{\sum_{i,j}E_{ij} r_{ij}}{\sum_{i,j} E_{ij}}.$$
 
Information from both layers is used to construct the
2 layer correlation variables. In the following
we differentiate between the data gathered in
the layer that recorded the cell with the maximum energy, the {\it
leading} layer, and  label them  with the superscript, $l$,
and the
data from the other layer, the {\it non-leading} layer,and label them
with the superscript, $n$.

The following 6 variables embody the 2 layer correlation information
used in the algorithms: the correlations  between  $E_{max}$ from one
layer and  $E_{4}$ from the other layer  are contained in the variables
 $C_{1}$ and $C_{2}$. Similarly, the
correlations  between $E_{4}$ and  $E_{8}$
are contained in the variables  $C_{3}$ and $C_{4}$ and, finally,
the correlations between $E_{8}$ and $E_{12}$ are contained in
 $C_{5}$ and $C_{6}$.
 
\begin{itemize}
\item $C_{1}$ is the difference, $E_{4}^{n} - E_{max}^{l}$,
 normalized to  $E_{total}$.
\item $C_{2}$ is  the difference, $E_{4}^{l} - E_{max}^{n}$,
normalized  to  $E_{total}$. Notice that  $C_{2}$ is similar to $C_{1}$
 with
the roles of the 2 layers interchanged.
\item $C_{3}$ is the difference
$E_{4}^{n} - E_{8}^{l}$  normalized to  $E_{total}$.
\item $C_{4}$ is the difference
$E_{4}^{l} - E_{8}^{n}$  normalized to  $E_{total}$.
 $C_{4}$ is similar to $C_{3}$ with
the roles of the 2 layers interchanged.

\item $C_{5}$ is the difference
$E_{8}^{n} - E_{12}^{l}$  normalized to  $E_{total}$.
\item $C_{6}$ is the difference
$E_{8}^{l} - E_{12}^{n}$  normalized to  $E_{total}$.
 $C_{6}$ is similar to $C_{5}$ with
the roles of the 2 layers interchanged.
\end{itemize}

\section{Analysis}
The study of the separation of  \mbox{$\gamma$/$\pi^{0}$}
showers is quantified by determining the rejection efficiency of
$\pi^{0}$ initiated showers as a function of the acceptance of
$\gamma$ initiated showers. Eight variables are used in these
analyses: the six correlations, $C_{1}$,.. ,$C_{6}$ and the two
average energy weighted radii, $<r>^l$ and $<r>^n$.
 
For every configuration of design parameters and
incident energy studied, two  samples of
  showers, one
initiated by $\g$'s and the other by $\pizero$'s, were generated and
propagated through the E.M. calorimeter and the SMD-dos.
The shower information collected by the SMD-dos was
used to construct the 8 variables previously described.
A program which uses the
Simulated Annealing Algorithm  was utilized
to delineate an 8-dimensional volume, corresponding to
the 8 variables used, which contains a fixed percentage of the $\g$
initiated showers while keeping to a minimum the number of $\pizero$
initiated showers. The delineation of this volume sets upper or lower
cutoff values for each of the 8 variables. The $\pizero$ rejection
efficiency is equal to the percentage of  $\pizero$
initiated showers which are excluded from the delineated
8-dimensional volume.
 
The shower information collected by the SMD was also processed using
the same analysis chain detailed above. However, since for the SMD
no correlation information is available, the following variables were
used: $E_{4}$, $E_{8}$, $E_{12}$ and $<r>$.
 
\begin{figure}[H]
\centering
\epsfig{file=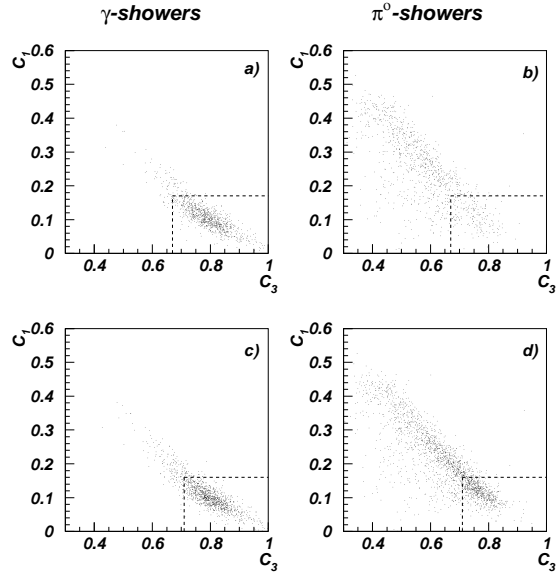,height=8.cm}
\caption{ Scatterplots of the correlation variables $C_{3}$ versus
$C_{1}$ for showers initiated by $\g$'s  and $\pizero$'s. a) 50 GeV
$\g$ showers and b) 50 GeV $\pizero$ showers. c) 100 GeV $\g$ showers
and  d) 100 GeV $\pizero$ showers. The respective cutoff values of
$C_{1}$ and $C_{3}$, for
90$\%$ $\g$ acceptance, at 50 GeV and 100 GeV, are shown by
dashed rectangles on the lower right corner of each figure.}
\label{cuts}
\end{figure}

The computer flow chart of the program used along with a short
description of those aspects of the program which are specific to
the  \mbox{$\gamma$/$\pi^{0}$} separation problem can be found
elsewhere \cite{telav}.
 In Fig.~\ref{cuts} are shown the projections,
for 50 and 100 GeV,  of the 8-dimensional volume of $\g$ initiated
showers and $\pizero$ initiated showers, separately,
unto the plane spanned by the
correlation variables $C_{1}$ and $C_{3}$.
The cutoff values of
$C_{1}$ and $C_{3}$ for
90$\%$ acceptance of the $\g$ initiated showers
define the rectangles shown in the lower right corners. For
50 GeV incident energy the
 $\pizero$ initiated showers overwhelmingly
  fall outside the acceptance rectangle giving high \rejpi  while for
100 GeV incident energy the number of
 $\pizero$ initiated showers excluded from the acceptance rectangle
is smaller hence giving  lower \rejpi.
 
The cutoff values for different incident energies are given in
Table~\ref{tabl1}.
These cutoff values as well as all other results presented in this paper,
for  the SMD-dos and SMD, were determined from samples of 500 showers
each which were generated under the following conditions:
\begin{itemize}
\item The distance of the E.M. calorimeter from the interaction
 region was fixed at 150 cm.
\item  The shower initiating particle, $\gamma$ or $\pi^{0}$, was
incident normally on the E.M. calorimeter. However, the direction of the
initiating particle was spread randomly within a small angular range
corresponding to an area of a few cells of the SMD-dos or SMD.
\item The \rejpi was calculated for $90\%$ $\g$ acceptance.
\item The SMD-dos or  SMD was positioned at a depth of
 6 $X_0$'s inside the  E.M. calorimeter.
\item Thickness of the scintillator layers set at 4 mm.
\item The nominal granularity, cell size, is $1 \times 1$ cm$^2$. In the
event that a different cell size was used, it is explicitly stated.
\end{itemize}
\begin{table}[H]
\begin{tabular}{|c||c|c|c|c|c|c||c|c|}
\hline
$E_{in}$ & $C_1 \leq$  & $C_2 \leq$  & $C_3 \geq$ & $C_4 \geq$ &
$C_5 \leq$ & $C_6 \leq$  & $\langle r \rangle^l \leq$ &
$\langle r \rangle^n \leq $  \\
\hline
30  & 0.18 & 0.70 & 0.65 & 0.21 & 0.65 & 0.17 & 0.58 & 0.80 \\
\hline
50  & 0.17 & 0.65 & 0.67 & 0.22 & 0.63 & 0.15 & 0.55 & 0.82 \\
\hline
80  & 0.17 & 0.68 & 0.70 & 0.20 & 0.62 & 0.16 & 0.53 & 0.85 \\
\hline
100 & 0.16 & 0.70 & 0.71 & 0.22 & 0.65 & 0.18 & 0.50 & 0.88 \\
\hline
120 & 0.15 & 0.67 & 0.71 & 0.23 & 0.60 & 0.14 & 0.50 & 0.90 \\
\hline
150 & 0.15 & 0.64 & 0.73 & 0.23 & 0.58 & 0.15 & 0.48 & 0.91 \\
\hline
\end{tabular}
\caption{Cutoff values of the six correlation variables,
$C_1$,.. ,$C_6$, and the two average energy weighted radii,
$<r>^l$ and $<r>^n$,  for incident energies of 30 to 150 GeV.}
\label{tabl1}
\end{table}
 
\section{Results and Discussion}
 
\begin{figure}[H]
\centering
\epsfig{file=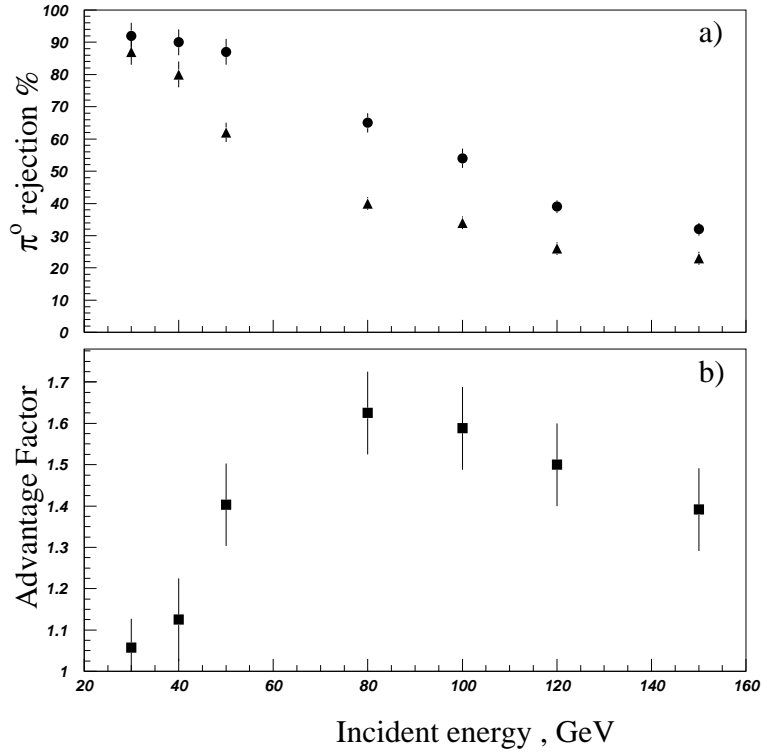,height=11.cm}
\caption{a)The $\pi^{0}$ rejection efficiency of the
SMD-dos ($\bullet$)  and a conventional
 SMD ($\blacktriangle$)  as a function of incident energy.
b)The `advantage factor' ($\blacksquare$)
 of the SMD-dos over the SMD as a function of
incident energy.}
\label{fig2}
\end{figure}

The $\pi^{0}$ rejection efficiency as a function of the incident
energy of the showering particle is shown in Fig.~\ref{fig2}a for
both the SMD-dos and SMD. Both detectors have equal granularity,
$1 \times 1$ cm$^2$ cells.
The $\pi^{0}$ rejection efficiency of the SMD-dos and the SMD decrease
as a function of energy;  however, the $\pi^{0}$ rejection efficiency
of the SMD-dos is higher than that of the SMD over the whole energy
range. The ratio of the rejection efficiency of the SMD-dos to
the SMD, the `advantage factor', is plotted as function of energy in
Fig.~\ref{fig2}b.
 
The value of the `advantage factor' is close to 1.0
at 30 GeV and it then rises to a maximum value of 1.63 at
80 GeV. With increasing energy the value of the `advantage
factor' decreases gradually reaching the value of 1.39 at 150 GeV,
the highest energy studied.
In Table~\ref{tabl2} are summarized the
the $\pi^{0}$ rejection efficiencies of the SMD-dos and
SMD, of equal granularity, for various incident energies. The `advantage
factor' has been calculated for each incident energy.
The following parameters, which are relevant to
\mbox{$\gamma$/$\pi^{0}$} separation problems are given, per energy,
in Table~\ref{tabl2}:
$\Theta_{12}$, the minimum opening angle in the lab system
of the  $\pi^{0} \to \gamma \gamma$ decay;
$d_{1,2}$, the distance between the two $\gamma$ centers at
the position of the SMD-dos or SMD  and
the ratio ${d_{1,2}/{\Delta}}$,
where $\Delta$ is the length of cell side, 1 cm.
 
\begin{table}[H]
\begin{tabular}{|c||c||c|c|c|c|c|} \hline
E$_{in}$ & $\pi^{0}$ Rejection &  $\Theta_{12}$ & $d_{1,2}$ &
$\underline{d_{1,2}}$  & $\pi^{0}$ Rejection  & `advantage  \\
$[GeV]$ & SMD-dos [$\%$] & $[mrad]$ & $[cm]$ & $\Delta$ & SMD [$\%$]
 & factor' \\
\hline
30  & 92$\pm$4 &  9.3 & 1.43  &  1.43  & 87$\pm$4   & 1.06$\pm$0.07  \\
\hline
40  & 90$\pm$4 &  7.0 & 1.07  &  1.07  & 80$\pm$4   & 1.13$\pm$0.08  \\
\hline
50  & 87$\pm$4 &  5.6 & 0.86 &  0.86  & 62$\pm$3    & 1.40$\pm$0.09 \\
\hline
80  & 65$\pm$3 &  3.5 & 0.55 &  0.55  & 40$\pm$2    & 1.63$\pm$0.11 \\
\hline
100 & 54$\pm$3 &  2.8 & 0.43 &  0.43  & 34$\pm$2    & 1.59$\pm$0.13 \\
\hline
120 & 39$\pm$2 &  2.3 & 0.36 &  0.36  & 26$\pm$2    & 1.50$\pm$0.14 \\
\hline
150 & 32$\pm$2 &  1.9 & 0.29 &  0.29  & 23$\pm$2    & 1.39$\pm$0.14 \\
\hline
\end{tabular}
\caption{The $\pi^{0}$ rejection efficiency of  the SMD-dos and
SMD, of equal granularity, is presented
for different values of incident energy.
The following parameters, which are explained in the text,
are given  per incident energy: $\Theta_{12}$,
$d_{1,2}$, the ratio ${d_{1,2}/{\Delta}}$ and the `advantage factor'.}
\label{tabl2}
\end{table}
 
\begin{figure}[H]
\centering
\epsfig{file=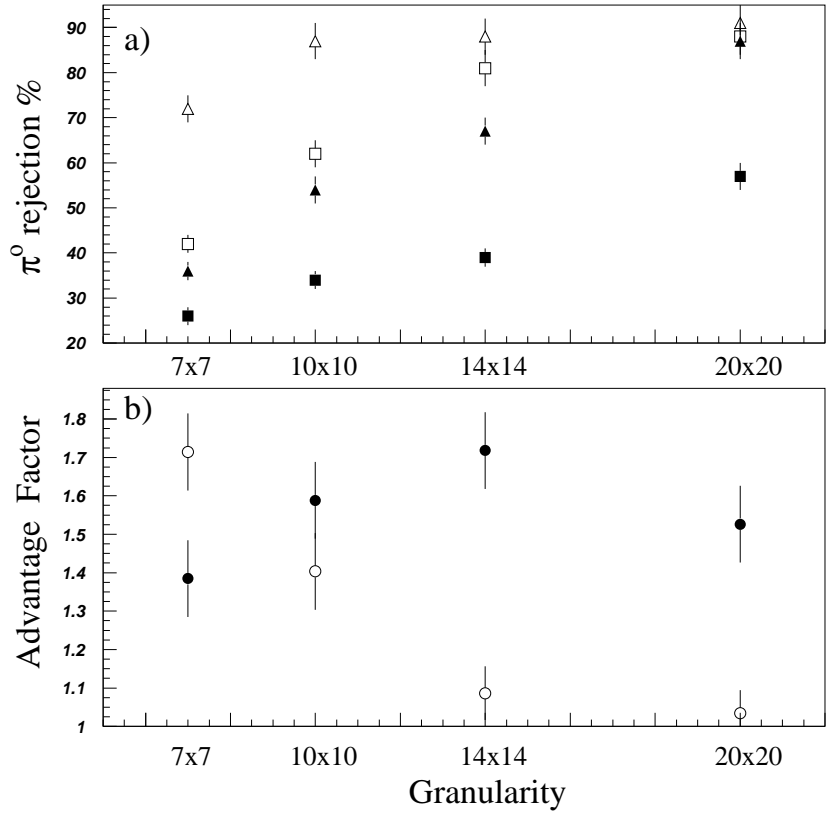,height=12.cm}
\caption{a)
The $\pi^{0}$ rejection efficiency, as a function of detector
granularity, of the SMD-dos ($\vartriangle$) and a conventional
 SMD ($\square$) at 50 GeV incident
energy; at 100 GeV, SMD-dos ($\blacktriangle$) and a conventional
 SMD ($\blacksquare$).
b) The `advantage factor' as a function of detector granularity at
50 GeV ($\circ$) and 100 GeV ($\bullet$).
The granularity is that of a
$10 \times 10$ cm$^2$ layer divided into $n \times n$ square cells.}
\label{fig3}
\end{figure}

The dependence of the $\pi^{0}$ rejection efficiency   of the
SMD-dos and SMD on the granularity of the detector
was studied at  50 and 100 GeV
incident energies. The results of studies
are shown in Figure~\ref{fig3} and are summarized in Table~\ref{tabl3}.
 The expected increase in
$\pi^{0}$ rejection efficiency with increasing  detector granularity
for both SMD-dos and SMD is seen in Fig.~5a.
 
For 50 GeV incident energy the $\pi^{0}$ rejection efficiency of the
SMD-dos and the SMD converge as the granularity increases.
The `advantage factor', plotted in Fig.~5b,
 is seen to fall rapidly from a value of
1.71 for granularity of $7 \times 7$ to a value of 1.09
for a granularity of $14 \times 14$. The `advantage factor'
for granularity of $20 \times 20$ is consistent with 1.0.
For 100 GeV incident energy the $\pi^{0}$ rejection efficiency
of both the SMD-dos and the SMD increase with increasing granularity;
however, the $\pi^{0}$ rejection efficiency of the SMD-dos is higher
over the whole energy range.
The `advantage factor'  has a value
of 1.4 for $7 \times 7$ granularity, it attains the
maximum value of 1.7 for granularity $14 \times 14$
and it then decreases to 1.5 for $20 \times 20$ granularity.

From the data given in Table~\ref{tabl3} it is possible to
compare
 the $\pi^0$ rejection efficiency performance of the SMD-dos
versus the SMD for the same number of channels. For 50 GeV incident
energy a  SMD-dos detector which has two $7 \times 7$ layers
needs 98 channels in order to read out its 98 cells while a SMD detector
of $10 \times 10$  granularity needs 100 channels. From
Table~\ref{tabl3} we determine that the SMD-dos is a factor of
1.16 better, 72$\%$ vs 62$\%$. A $10 \times 10$  SMD-dos
is better than a $14 \times 14$  SMD by a factor of 1.07.
For 100 GeV incident energy we find that
a $10 \times 10$  SMD-dos is superior to a
 $14 \times 14$  SMD by a factor of 1.38. Also a
 $14 \times 14$  SMD-dos is a factor of 1.18 better than a
 $20 \times 20$  SMD.
\begin{table}[H]
\begin{tabular}{|c|c|c|c|c|c|}
\hline
\multicolumn{6}{|c|}{ Energy = 50GeV}\\
\hline
Number & $\Delta$ & $\underline{d_{1,2}}$ & $\pi^{0}$Rejection &
 $\pi^{0}$ Rejection & `advantage\\
of cells & $[cm]$ & $\Delta$ & SMD-dos [$\%$] & SMD [$\%$] & factor'  \\
\hline
7$\times$7   & 1.43 & 0.60   & 72$\pm$3 & 42$\pm$2 & 1.71$\pm$0.11 \\
\hline
10$\times$10 & 1.0  & 0.86   & 87$\pm$4 & 62$\pm$3 & 1.40$\pm$0.09 \\
\hline
14$\times$14 & 0.72 & 1.20   & 88$\pm$4 & 81$\pm$4 & 1.09$\pm$0.07 \\
\hline
20$\times$20 & 0.50 & 1.72   & 91$\pm$4 & 88$\pm$4 & 1.03$\pm$0.07 \\
\hline
\hline
\multicolumn{6}{|c|}{ Energy = 100 GeV}\\
\hline
Number   & $\Delta$ & $\underline{d_{1,2}}$ & $\pi^{0}$ Rejection &
 $\pi^{0}$ Rejection & `advantage\\
of cells & $[cm]$ & $\Delta$ & SMD-dos [$\%$] & SMD [$\%$] & factor'  \\
\hline
7$\times$7    &  1.43 & 0.30  &  36$\pm$2  & 26$\pm$2   & 1.38$\pm$0.13 \\
\hline
10$\times$10  &  1.0  & 0.43  &  54$\pm$2  & 34$\pm$3   & 1.59$\pm$0.13 \\
\hline
14$\times$14  &  0.72 & 0.60  &  67$\pm$2  & 39$\pm$3   & 1.71$\pm$0.12 \\
\hline
20$\times$20  &  0.5  & 0.86  &  87$\pm$3  & 57$\pm$4   & 1.52$\pm$0.11 \\
\hline
\end{tabular}
\caption{
The $\pi^{0}$ rejection efficiencies of the SMD-dos and SMD, the
`advantage factor's,
the values of $\Delta$ and the ratio $d_{1,2}/ \Delta$, are given
for different detector granularities, for 50
and 100 GeV incident energies.}
\label{tabl3}
\end{table}
The advantage of using 2 layers which are diagonally off-set with
respect to each other as oppossed to
 using a single layer, given that the total number of cells of each
detector is the same,
can be understood as follows:
the
\mbox{$\gamma$/$\pi^{0}$} separation power of a granular
detector is strongly dependent on the distance of the shower center
from an intersection. The success rate of any available
 \mbox{$\gamma$/$\pi^{0}$} separation
algorithm is higher when the deposited shower energy is distributed
among many cells. As previously stated, if all the deposited energy
is collected by a single cell it is impossible to determine the
identity of the showering particle.
The distance between intersections for a SMD-dos
is smaller by a factor of $1/\sqrt{2}$ than that of a SMD when both
detectors have equal number of cells. Therefore, a showering particle
incident on a SMD-dos detector will, on the average, have its shower
center nearer to an intersection than if it were incident on a
SMD detector.
\begin{figure}[H]
\centering
\epsfig{file=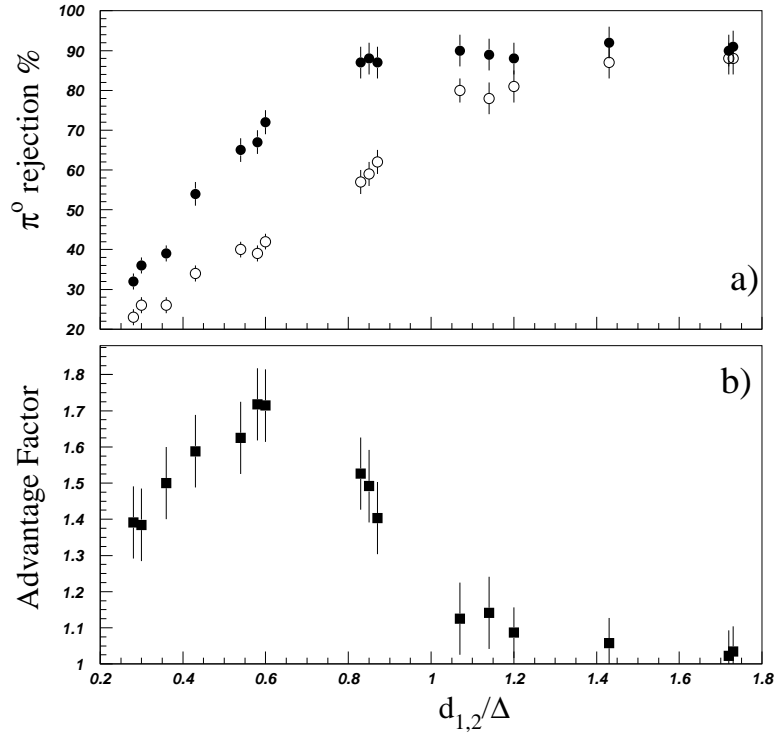,height=11.cm}
\caption{a) The $\pi^{0}$ rejection efficiency of the SMD-dos
($\bullet$) and a conventional SMD ($\circ$) as a function of the
ratio,  $d_{1,2}/ \Delta$.
b) The `advantage factor' as a function of the ratio,
$d_{1,2}/ \Delta$.}
\label{fig5}
\end{figure}
The ratio of the 2 variables, $d_{1,2}$, the distance between the
two $\gamma$ centers at the position of the SMD-dos or SMD and
$\Delta$, the length of cell side, ${d_{1,2}/{\Delta}}$, is a useful
a measure of the \mbox{$\gamma$/$\pi^{0}$} separation difficulty.
The
 \mbox{$\gamma$/$\pi^{0}$} separation becomes more difficulty as the
ratio ${d_{1,2}/{\Delta}}$ tends to small values while it is
significantly easier as the ratio tends to values above 1.0.
The behaviour of the
 $\pi^{0}$ rejection efficiency and the `advantage factor' as a
function of $d_{1,2}/ \Delta$ are shown in Figure~\ref{fig5}. The
$\pi^{0}$ rejection efficiency, as expected, increases for both the
SMD-dos and the SMD as the value of this ratio increases. We notice
that the `advantage factor' has an appreciable value only
in the range of from $\sim$0.2 to $\sim$1. Hence we find that
when the
 \mbox{$\gamma$/$\pi^{0}$} separation is very difficulty, values
smaller than $\sim$0.2, than the efficiencies of the SMD-dos and
SMD are comparable and likewise when the separation becomes easy,
values higher than $\sim$1.0, than again no advantage of SMD-dos
over SMD.
 
If we express, $d_{1,2}$ in terms of the incident energy, $E_{inc}$,
 mass of the $\pi^0$ and, $L$, distance
to SMD-dos or SMD from the interaction point, then we can write the
following expression:

$$\frac{d_{1,2}}{\Delta}=\frac{2\times m_{\pi}\times L}{E_{inc}\times \Delta}.$$
 
Hence the ratio, $d_{1,2}/ \Delta$, is a function of 3 independent
variables,
$\Delta$, $L$ and ${E_{inc}}$. So for instance if
 we fix the value of 2 of these
variables we can read off from   Figure~\ref{fig5} the
 permissible range of the third variable, for a desired
 $\pi^{0}$ rejection efficiency.
Figure~\ref{fig5} can be used, in general, in order to get the
 $\pi^{0}$ rejection efficiency and `advantage factor' for different
combinations of the 3 variables not directly studied in this paper.
However, it would be prudent to limit the use of these figures to
values of the 3 variables which are within a factor of 2 of the
ranges explicitly studied in this paper.
 
\section{Conclusions}
 
We have designed and extensively studied
a new geometry SMD consisting of two layers of scintillator
which are diagonally off-set with respect to each other. The layers
of scintillator are subdivided into square cells. This geometry has the
virtue of increasing substantially the number of intersections,
per unit area, seen by the incoming shower.
A comparison of the SMD-dos with a
conventional geometry single-layer and mutiple-layer SMD, having the
same granularity, shows that the SMD-dos yields higher
$\pi^{0}$ rejection efficiencies.
 For 90$\%$ $\g$ acceptance the SMD-dos yields
\rejpis  of  92$\pm$4 $\%$, 87$\pm$4 $\%$ and 32$\pm$2 $\%$,
respectively, for 30, 50 and 150 GeV incident energies.
 We find that the SMD-dos is
superior to the SMD, of equal granularity, by an average factor
of $\sim$ 1.5 over the 50 to 150 GeV energy range.
We also find that the SMD-dos gives better $\pi^0$ rejection,
 for the same number of channels, than a SMD. At 100 GeV a
$10 \times 10$ cells per layer SMD-dos, which utilizes 200 channels,
is a factor of  $\sim$ 1.4 better than a
$14 \times 14$ cells, single layer SMD, which utilizes an equal number of
channels.
The $\pi^{0}$ rejection efficiency of the SMD-dos and SMD and the
`advantage factor' are plotted as a function of the
ratio,  $d_{1,2}/ \Delta$. With the help of these plots it is possible
to investigate the relative importance of the 3 variables,
$\Delta$, $L$ and ${E_{inc}}$,
over a range which goes beyond the work presented in this paper.
 
At hadron - hadron colliders the signature of choice for the
detection of the Higgs particle, in the mass range of 120 to 160 GeV,
is via the decay \mbox{$H \to \gamma \gamma$}. The addition of a
SMD-dos to the planned detectors at the LHC would significantly
reduce the background to the $\gamma$ signal coming from prolific
$\pi^0$ production.

\end{document}